
\documentstyle{amsppt}

\NoBlackBoxes
\magnification=\magstep1
\hsize=6.5truein
\vsize=8.5truein
\document
\baselineskip=.15truein

\topmatter
\title   The $[n_1,n_2,\ldots,n_s]$--th reduced KP hierarchy and $W_{1+\infty}$
Constraints
\endtitle
\author
Johan van de Leur
\endauthor
\thanks
E-mail: vdleur\@math.ruu.nl
\endthanks
\abstract
To every partition $n=n_1+n_2+\cdots+n_s$ one can associate a vertex
operator realization of the Lie algebras $a_{\infty}$ and $\hat{gl}_n$.
Using this construction  we obtain reductions of the $s$--component KP
hierarchy,
reductions which are related to these partitions.
In this way we obtain matrix KdV type
equations. We show that the following two constraints on a KP $\tau$--function
are equivalent (1) $\tau$ is a $\tau$--function of the
$[n_1,n_2,\ldots,n_s]$--th reduced KP hierarchy which satisfies string
equation, $L_{-1}\tau=0$,
(2) $\tau$  satisfies the vacuum constraints of the
$W_{1+\infty}$ algebra.
\endabstract
\endtopmatter
\vskip 10pt

\noindent The results of this paper were presented at the V International
Conference on Mathematical Physics, String Theory and
Quantum Gravity at Alushta, June 10-20 1994 and at the Oberwolfach conference
on Integrable Systems from a Quantum Point of View, June 19-25 1994. A longer
edition of this paper [V], containing more details and proofs, will appear
elsewhere.

\vskip 10pt
\subheading{\S 1.  $a_{\infty}$ and the KP hierarchy in
the fermionic picture [KV]}

\vskip 10pt
\noindent
Consider the infinite dimensional complex  Lie algebra
$a_{\infty} := \overline{gl_{\infty}}\bigoplus {\Bbb C}c$, where
$$\overline{gl_{\infty}} = \{ a = (a_{ij})_{i,j \in {\Bbb Z}+\frac{1}{2}}
|\  a_{ij}=0\  \text{if}\   |i-j|>>0\},$$
with Lie bracket defined by
$$[a+\alpha c,b+\beta c]=ab-ba+\mu (a,b)c , \tag{1.1}$$
for $a,b\in \overline{gl_{\infty}}$ and $\alpha ,\beta\in{\Bbb C}$.
Here $\mu$ is the following 2--cocycle:
$$\mu (E_{ij},E_{kl})=\delta_{il}\delta_{jk}(\theta(i)-\theta(j)),
\tag{1.2}$$
where $E_{ij}$ is the matrix with a $1$ on the $(i,j)$-th entry and zeros
elsewhere and $\theta :{\Bbb R}\to {\Bbb C}$ is the step--function  defined by
$$\theta(i):=\cases 0 & \text{if}\ i>0,\\
                    1 & \text{if}\ i\le 0.\endcases \tag{1.3}$$
Let ${\Bbb C}^{\infty} = \bigoplus_{j \in {\Bbb Z}+\frac{1}{2}} {\Bbb
C} v_{j}$ be the infinite dimensional complex vector space with fixed
basis $\{ v_{j}\}_{j \in {\Bbb Z}+\frac{1}{2}}$. The Lie algebra $a_{\infty}$
acts linearly on
${\Bbb C}^{\infty}$ via the usual formula:
$$E_{ij} (v_{k}) = \delta_{jk} v_{i}.$$

Let ${\Cal C}\ell$ be the Clifford algebra with generators
$\psi^{\lambda}_{i}$,
$i \in {\Bbb Z}+\frac{1}{2}, \lambda ,\mu = +,-$, satisfying the following
relations:
$$\psi^{\lambda}_{i} \psi^{\mu}_{j} + \psi^{\mu}_{j}
\psi^{\lambda}_{i} = \delta_{\lambda ,-\mu} \delta_{i,-j}. \tag{1.4}$$
We define an irreducible  ${\Cal C}\ell$--module $F$ by introducing a vacuum
vector $|0\rangle$ such that
$$\psi^{\pm}_{j} |0\rangle = 0 \ \text{for}\ j > 0 . \tag{1.5}$$
Define  a representation $\hat r$ of $a_{\infty}$ on $F$ by
$$ \hat r(E_{ij})=:\psi^-_{-i}\psi^+_j:,\quad \hat r(c)=I,
$$
where $:\ :$ stands for the {\it normal ordered product} defined in
the usual way $(\lambda ,\mu = +$ or $-$):
$$:\psi^{\lambda (i)}_{k} \psi^{\mu (j)}_{\ell}: = \cases \psi^{
\lambda (i)}_{k}
\psi^{\mu (j)}_{\ell}\ &\text{if}\ \ell \ge k \\
-\psi^{\mu (j)}_{\ell} \psi^{\lambda (i)}_{k} &\text{if}\ \ell <
k.\endcases \tag{1.6}$$

Define the {\it charge decomposition}
$$F = \bigoplus_{m \in {\Bbb Z}} F^{(m)} \tag{1.7}$$
by letting

$$\text{charge}(|0\rangle ) = 0\ \text{and charge} (\psi^{\pm}_{j}) =
\pm 1. \tag{1.8}$$
It  is easy to see that each $F^{(m)}$ is irreducible with
respect to  $a_{\infty}$.

We are now able to define the {\it KP hierarchy in the fermionic
picture}, it is the equation
$$\sum_{k \in {\Bbb Z}+\frac{1}{2}} \psi^{+}_{k} \tau \otimes \psi^{-}_{-k}
\tau = 0, \tag{1.9}$$
for
$\tau\in F^{(0)}$. One can prove (see e. g. [KP2] or [KR]) that  this equation
characterizes  the group orbit of the vacuum vector  $|0\rangle$  for the  the
group $GL_{\infty}$. Since the group does not play an important role in this
paper, we will not introduce it here.
\vskip 10pt
\subheading{\S 2.  Vertex operator constructions}

\vskip 10pt
\noindent We will now sketch how one can  construct vertex realizations of
the affine Lie algebra $\hat{gl}_n$,
following [TV] (see also [KP1] and [L]). From now on let $n=n_1+
n_2+\cdots+n_s$ be a partition of $n$ into $s$ parts,  and denote by
$N_a=n_1+n_2+\cdots +n_{a-1}$.
We begin by relabeling the basis vectors $v_j$ and with them the corresponding
fermionic  operators:
($1\le a \le s,\ 1\le p\le n_a,\ j\in {\Bbb Z}$)
$$\align
v^{(a)}_{n_aj-p+\frac{1}{2}}&=v_{nj-N_a-p+\frac{1}{2}} \\
\psi^{\pm (a)}_{n_a j \mp p\pm \frac{1}{2}}&=\psi^{\pm }_{n_aj\mp N_a\mp
p\pm\frac{1}{2}}
\tag{2.1}
\endalign$$
Notice that with this relabeling we have:
$\psi^{\pm (a)}_{k}|0\rangle = 0\ \text{for}\ k > 0.$
We also rewrite the $E_{ij}$'s:
$$E^{(ab)}_{n_aj-p+\frac{1}{2},n_bk-q+\frac{1}{2}}=E_{nj-N_a-p+\frac{1}{2},nk-N_b-q+\frac{1}{2}},\tag{2.2}$$
then  $\hat r(E^{(ab)}_{jk})=:\psi^{-(a)}_{-j}\psi^{+b}_k:$.

Introduce the  fermionic fields $(z \in {\Bbb C}^{\times})$:
$$\psi^{\pm (a)}(z) \overset{\text{def}}\to{=} \sum_{k \in {\Bbb
Z}+\frac{1}{2}} \psi^{\pm
(a)}_{k} z^{-k-\frac{1}{2}}.\tag{2.3}$$
Let $N$ be the least common multiple of $n_1,n_2,\ldots,n_s$.
It was shown in [TV] that the modes of the fields
$$:\psi^{+(a)}(\omega_a^p z^{N\over n_a})\psi^{-(b)}(\omega_b^q z^{N\over
n_b}):, \tag{2.4}$$
for $1\le a,b\le s$, $1\le p\le n_a$, $1\le q\le n_b$, where $\omega_a=e^{2\pi
i/n_a}$, together with the identity,
generate a representation of $\hat{gl}_n$ with center $K=1$.
It is easy to see that restricted to $\hat{gl}_n$,
$F^{(0)}$ is its basic highest weight representation (see [K, Chapter
12]).
{}From (2.4) it  is obvious, that if we obtain vertex operators for the
fermionic fields
$\psi^{\pm (a)}(z)$, we also have a vertex operator realization of
$\hat{gl}_n$.

Next we introduce special bosonic fields ($1\le a\le s$):
$$\alpha^{(a)}(z) \equiv \sum_{k \in {\Bbb Z}} \alpha^{(a)}_{k} z^{-k-1}
\overset{def}\to{=} :\psi^{+(a)}(z) \psi^{-(a)}(z):. \tag{2.5}$$
The operators $\alpha^{(a)}_{k} $
satsify the canonical commutation relation of the associative
oscillator algebra,  which we
denote by ${\frak a}$:
$$[\alpha^{(i)}_{k},\alpha^{(j)}_{\ell}] =
k\delta_{ij}\delta_{k,-\ell},\tag{2.6}$$
and one has
$$\alpha^{(i)}_{k}|m\rangle = 0 \ \text{for}\ k > 0.$$

This realization  of $\hat{gl}_n$,  has a natural Virasoro algebra. In [TV], it
was shown that the following two sets of operators have the same action on $F$.
$$\align
&L_k=\sum_{i=1}^s\{ \sum_{j\in{\Bbb Z}}{1\over
2n_i}:\alpha_{-j}^{(i)}\alpha_{j+n_ik}^{(i)}:+\delta_{k0}{{n_i^2-1}\over
24n_i}\},\tag{2.7a} \\
&H_k=\sum_{i=1}^s\{\sum_{j\in{\Bbb Z}+\frac{1}{2}}({j\over n_i}+{k\over
2}):\psi^{+(i)}_{-j}\psi^{-(i)}_{j+n_ik}:+\delta_{k0}{{n_i^2-1}\over
24n_i}\}. \tag{2.7b} \endalign
$$
So $L_k=H_k$,
$$[L_k,\psi_j^{\pm (i)}]=-(\frac{j}{n_i}+\frac{k}{2})\psi_{j+n_ik}^{\pm(i)}
\tag{2.8}
$$
 and
$$[L_k,L_\ell ]=(k-\ell)L_{k+\ell}+\delta_{k,-\ell}{{k^3-k}\over 12}n. $$

We will now
describe the $s$-component boson-fermion
correspondence (see [KV]).  Let ${\Bbb C}[x]$ be the space of polynomials in
indeterminates $x = \{ x^{(i)}_{k}\},\ k = 1,2,\ldots ,\ i =
1,2,\ldots ,s$.  Let $L$ be a lattice with a basis $\delta_{1},\ldots
,\delta_{s}$ over ${\Bbb Z}$ and the symmetric bilinear form
$(\delta_{i}|\delta_{j}) = \delta_{ij}$, where $\delta_{ij}$ is the
Kronecker symbol.  Let
$$\varepsilon_{ij} = \cases -1 &\text{if $i > j$} \\
1 &\text{if $i \leq j$.} \endcases \tag{2.9}$$
Define a bimultiplicative function $\varepsilon :\ L \times L @>>> \{
\pm 1 \}$ by letting
$$\varepsilon (\delta_{i}, \delta_{j}) = \varepsilon_{ij}.
\tag{2.10}$$
Let $\delta = \delta_{1} + \ldots + \delta_{s},\  Q= \{ \gamma \in
L|\ (\delta | \gamma ) = 0\}$, $\Delta = \{ \alpha_{ij} :=
\delta_{i}-\delta_{j}| i,j = 1,\ldots ,s,\ i \neq j \}$.  Of course
$Q$ is the root lattice of $sl_{s}({\Bbb C})$, the set $\Delta$
being the root system.

Consider the vector space ${\Bbb C}[L]$ with basis $e^{\gamma}$,\
$\gamma \in L$, and the following twisted group algebra product:
$$e^{\alpha}e^{\beta} = \varepsilon (\alpha ,\beta)e^{\alpha +
\beta}. \tag{2.11}$$
Let $B = {\Bbb C}[x] \otimes_{\Bbb C} {\Bbb C}[L]$ be the tensor
product of algebras.  Then the $s$-component boson-fermion
correspondence is the vector space isomorphism
$$\sigma :F @>\sim >> B, \tag{2.12}$$
given by $\sigma (|0\rangle )=1$ and
$$\sigma\psi^{\pm (a)}(z) \sigma^{-1}=e^{\pm \delta_a}z^{\pm \delta_a}
\exp(\pm\sum_{k=1}^{\infty}x_k^{(a)}z^k)\exp(\mp\sum_{k=1}^{\infty}
{\partial\over \partial x_k^{(a)}}{z^{-k}\over k}),\tag {2.13}$$
where
$$\delta_a (p(x) \otimes e^{\gamma}) =
(\delta_{a}|\gamma ) p(x) \otimes e^{\gamma}.\tag{2.14}$$
The transported charge then is as follows:
$$
\text{charge}(p(x)\otimes e^{\gamma}) = (\delta |\gamma).
$$
We denote the transported charge decomposition by
$$B = \bigoplus_{m \in {\Bbb Z}} B^{(m)}.$$
The transported action of the operators $\alpha^{(i)}_{m}$ is given by
$$\cases
\sigma \alpha^{(j)}_{-m}\sigma^{-1}(p(x) \otimes e^{\gamma}) =
mx^{(j)}_{m}p(x)\otimes e^{\gamma},\ \text{if}\ m > 0, &\  \\
\sigma \alpha^{(j)}_{m} \sigma^{-1}(p(x) \otimes e^{\gamma}) = \frac{\partial
p(x)}{\partial x_{m}} \otimes e^{\gamma},\ \text{if}\ m > 0, &\  \\
\sigma \alpha^{(j)}_{0} \sigma^{-1} (p(x) \otimes e^{\gamma}) =
(\delta_{j}|\gamma ) p(x) \otimes e^{\gamma}
. & \
\endcases \tag{2.15}
$$

If one substitutes (2.13) into (2.4), one obtains the vertex operator
realization of $\hat{gl}_n$ which is related to the partition
$n=n_1+n_2+\cdots+n_s$ (see [TV] for more details).

Using the isomorphism $\sigma$ we can reformulate the KP hierarchy
(1.9) in the bosonic picture.

We start by observing that (1.9) can be rewritten as follows:
$$\text{Res}_{z=0}\ dz ( \sum^{s}_{j=1} \psi^{+(j)}(z)\tau
\otimes \psi^{-(j)}(z)\tau ) = 0,\ \tau \in F^{(0)}.
\tag{2.16}$$
Notice that for $\tau \in F^{(0)},\ \sigma (\tau) = \sum_{\gamma \in Q}
\tau_{\gamma}(x)e^{\gamma}$.
 Here and further  we write $\tau_{\gamma}(x)e^{\gamma}$ for
$\tau_{\gamma} \otimes
e^{\gamma}$.  Using  (2.13), equation (2.16) turns under $\sigma
\otimes \sigma :\ F \otimes F \overset\,\,\sim\to\longrightarrow
{\Bbb C}[x^{\prime},x^{\prime \prime}]
\otimes ({\Bbb C}[L^{\prime}] \otimes {\Bbb C}[L^{\prime \prime}])$ into the
following set of equations;
 for all $\alpha ,\beta \in L$ such that $(\alpha
|\delta ) = -(\beta |\delta ) = 1$ we have:
$$\aligned
&\text{Res}_{z=0} ( dz
 \sum^{s}_{j=1} \varepsilon (\delta_{j}, \alpha-\beta)
z^{(\delta_{j}|\alpha - \beta - 2\delta_{j})}  \\
 &\times \exp
(\sum^{\infty}_{k=1} (x^{(j)^{\prime}}_{k} - x^{(j)^{\prime
\prime}}_{k})z^{k})
\exp (-\sum^{\infty}_{k=1} (\frac{\partial}{\partial x^{(j)^{\prime}}_{k}}
 - \frac {\partial}{\partial x^{(j)^{\prime
\prime}}_{k}})\frac{z^{-k}}{k}) \\
& \tau_{\alpha -
\delta_{j}}(x^{\prime})(e^{\alpha})^{\prime}\tau_{\beta + \delta_{j}}(x^{\prime
\prime})(e^{\beta})^{\prime\prime})
= 0  . \endaligned \tag{2.17}$$

\vskip 10pt
\subheading{\S 3. The algebra of formal pseudo-differential operators and the
$s$-component KP hierarchy as a dynamical system [KV]}

\vskip 10pt
\noindent We proceed now to rewrite the formulation (2.17) of the
$s$-component KP hierarchy in terms of formal pseudo-differential
operators, generalizing the results of [DJKM] and [JM].
For each $\alpha \in \ \text{supp}\ \tau :=\{\alpha\in Q| \tau=\sum_{\alpha\in
Q}\tau_{\alpha}e^{\alpha}, \tau_{\alpha}\ne 0\}$ we define the (matrix
valued) functions
$$V^{\pm} (\alpha ,x,z) = (V^{\pm}_{ij}(\alpha ,x,z))^{s}_{i,j=1}
\tag{3.1}$$
as follows:
$$\aligned
&V^{\pm}_{ij}(\alpha ,x,z) \overset{\text{def}}\to{=}
\varepsilon (\delta_{j} , \alpha + \delta_{i})
 z^{(\delta_{j}|\pm \alpha + \delta_{i}-\delta_{j})} \\
& \times \exp (\pm \sum^{\infty}_{k=1} x^{(j)}_{k} z^{k})
\exp(\mp \sum^{\infty}_{k=1} \frac{\partial}{\partial
x^{(j)}_{k}} \frac{z^{-k}}{k}) \tau_{\alpha  \pm
(\delta_{i}-\delta_{j})}  (x)/\tau_{\alpha}(x) .
\endaligned \tag{3.2}
$$
It is easy to see that equation (2.17) is equivalent to the
following bilinear identity:
$$Res_{z=0}V^{+}(\alpha ,x,z)\ ^{t}V^{-}(\beta ,x^{\prime},z)dz = 0\
\text{for all}\ \alpha ,\beta \in Q. \tag{3.3}$$
Define $s \times s$ matrices $W^{\pm (m)} (\alpha ,x)$ by the
following generating series (cf. (3.2)):
$$
\sum^{\infty}_{m=0}
W^{\pm (m)}_{ij} (\alpha ,x)(\pm z)^{-m}
= \varepsilon_{ji}z^{\delta_{ij}-1} (\exp \mp
\sum^{\infty}_{k=1} \frac{\partial}{\partial x^{(j)}_{k}}\frac{z^{-k}}{k})
\tau_{\alpha \pm
(\delta_i-\delta_j)} (x))/\tau_{\alpha} (x). \tag{3.4}
$$
We see from (3.2) that $V^{\pm}(\alpha ,x,z)$ can be written in the
following form:
$$V^{\pm}(\alpha ,x,z) = (\sum^{\infty}_{m=0}
W^{\pm (m)}(\alpha ,x)R^{\pm}(\alpha ,\pm z)(\pm
z)^{-m})e^{\pm z \cdot x}, \tag{3.5}$$
where
$$z \cdot x^{(j)} = \sum^{\infty}_{k=1} x^{(j)}_{k} z^{k},\ e^{z
\cdot x} = diag (e^{z\cdot x^{(1)}} ,\ldots ,e^{z\cdot x^{(s)}} )$$
and
$$R^{\pm}(\alpha ,z) = \sum^{s}_{i=1}
\varepsilon (\delta_{i}, \alpha ) E_{ii} (\pm z)^{\pm
(\delta_{i}|\alpha )}. \tag{3.6}$$
Here and further $E_{ij}$ stands for the $s \times s$ matrix whose
$(i,j)$ entry is $1$ and all other entries are zero. Let
$$\partial = \frac{\partial}{\partial x^{(1)}_{1}} + \ldots +
\frac{\partial}{\partial x^{(s)}_{1}},$$ we can now rewrite
$V^{\pm}(\alpha ,x,z)$ in
terms of formal pseudo-differential operators
$$P^{\pm}(\alpha ) \equiv P^{\pm} (\alpha ,x,\partial ) =
I_{s} + \sum^{\infty}_{m=1} W^{\pm (m)} (\alpha ,x)\partial^{-m}\
\text{and}\ R^{\pm}(\alpha ) = R^{\pm}(\alpha ,\partial) \tag{3.7}$$
as follows:
$$V^{\pm}(\alpha ,x,z) = P^{\pm } (\alpha )
R^{\pm}(\alpha)e^{\pm z \cdot x}. \tag{3.8}$$

As usual one denotes the differential part of $P(x,\partial )$ by
$P_{+}(x,\partial ) = \sum^{N}_{j=0} P_{j}(x) \partial^{j},$
and writes $P_{-} = P-P_{+}$. The  linear  anti-involution   $*$ is defined by
the
following formula:
$$(\sum_{j} P_{j}\partial^{j})^{*} = \sum_{j} (-\partial )^{j}
\circ ^{t}\! P_{j}. \tag{3.9}$$
Here and further $^{t}P$ stands for the transpose of the matrix $P$.
Then one has the following fundamental lemma  (see [KV]):

\proclaim{Lemma 3.1} If $P,Q \in \Psi$ are such that
$$Res_{z=0} (P(x,\partial) e^{z\cdot x})  \ ^{t}
(Q(x^{\prime},\partial^{\prime}) e^{-z\cdot x^{\prime}})  dz = 0,
$$
then $(P \circ Q^{*})_{-} = 0$.
\endproclaim

Using this Lemma, Victor Kac and the author showed in [KV] that
given $\beta \in \text{supp}\ \tau$, all the
pseudo-differential operators $P^{\pm} (\alpha )$, $\alpha \in \text{supp}\
\tau$, are
completely determined by $P^+(\beta )$ from the following equations
$$\align
&R^{-}(\alpha ,\partial)^{-1} = R^{+}(\alpha
,\partial)^{*} ,
\tag{3.10}\\
&P^{-}(\alpha  ) = (P^{+}(\alpha  )^{*})^{-1} ,
\tag{3.11}\\
&(P^{+}(\alpha)R^{+}(\alpha - \beta)P^{+}(\beta)^{-1})_{-} = 0\
\text{for all}\ \alpha ,\beta \in \text{supp}\ \tau . \tag{3.12}
\endalign$$
They also showed
 the following

\proclaim{Proposition 3.2}  Consider $V^{+}(\alpha ,x,z)$ and $V^{-}(\alpha
,x,z),$ \newline $\alpha \in Q$, of the
form (3.8), where $R^{\pm}(\alpha ,z)$ are given by (3.6).  Then the
bilinear identity (3.3) for all $\alpha ,\beta \in \ \text{supp}\
\tau$ is equivalent to the Sato equation:
$$
\frac{\partial P}{\partial x^{(j)}_{k}} = -(PE_{jj} \circ
\partial^{k} \circ P^{-1})_{-} \circ P. \tag{3.13}$$
for each $P =
P^{+}(\alpha )$ and the matching conditions (3.10-12)  for
all $\alpha ,\beta \in \ \text{supp} \ \tau$.
\endproclaim

  Fix $\alpha \in Q$, introduce the following formal pseudo-differential
operators $L(\alpha),\  C^{(j)}(\alpha )$, and
differential operators  $B^{(j)}_{m}(\alpha)$:
$$\aligned
L \equiv L(\alpha )
  & = P^{+}(\alpha) \circ \partial \circ P^{+}(\alpha)^{-1}, \\
C^{(j)} \equiv C^{(j)}(\alpha ) &=
P^{+}(\alpha)E_{jj} P^{+}(\alpha)^{-1}, \\
B^{(j)}_{m} \equiv B^{(j)}_{m}(\alpha) &=
(P^{+}(\alpha)E_{jj} \circ \partial^{m} \circ
P^{+}(\alpha)^{-1})_{+}.
\endaligned \tag{3.14}
$$
Then
$$
\sum^{s}_{i=1} C^{(i)} = I_{s}, \
C^{(i)}L = LC^{(i)},\ C^{(i)}C^{(j)} = \delta_{ij} C^{(i)}.
 \tag{3.15}$$
\proclaim{Proposition 3.3} If for every $\alpha \in Q$ the
formal pseudo-differential operators $L \equiv L(\alpha)$ and $C^{(j)} \equiv
C^{(j)}(\alpha)$ of the form (3.14) satisfy conditions (3.15) and if the
equations
$$
\cases LP = P\partial &\ \\
C^{(i)}P = PE_{ii} &\ \\
\displaystyle{\frac{\partial P}{\partial x^{(i)}_{k}} = -(L^{(i)k})_{-} P,\
\text{where}\  L^{(i)} = C^{(i)}L.}
\endcases
\tag{3.16}$$
 have a solution $P \equiv P^+(\alpha)$ of the form (3.7), then the
differential operators $B^{(j)}_{k} \equiv B^{(j)}_{k}(\alpha)$
satisfies the following
conditions:
$$\left\{ \matrix \displaystyle{\frac{\partial L}{\partial x^{(j)}_{k}} =
[B^{(j)}_{k},L],}  & \   \\
\displaystyle{\frac{\partial C^{(i)}}{\partial x^{(j)}_{k}} =
[B^{(j)}_{k},C^{(i)}],} & \
\  \endmatrix \right. \tag{3.17}$$
\endproclaim
Finally, we introduce one more pseudo--differential operator
$$M(\alpha):=P(\alpha)R(\alpha)\sum_{a=1}^s\sum_{k=1}^{\infty} kx_k^{(a)}
\partial^{k-1}E_{aa}R(\alpha)^{-1}P(\alpha)^{-1}.\tag{3.18}$$
Then
$$M(\alpha) V^+(\alpha,x,z)={{\partial V^+(\alpha,x,z)}\over \partial
z}\tag{3.19}$$
and  one easily checks that $[L(\alpha),M(\alpha)]=1$

\vskip 10pt
\subheading{\S4 . $[n_1,n_2,\ldots ,n_s]$-reductions of the $s$-component KP
hierarchy}
\vskip 10pt
\noindent If one restricts the, to the partition $n=n_1+n_2+\cdots +n_s$
related, vertex
operator construction of $\hat{gl}_n$ in the vector space $B^{(0)}$ to
$\hat{sl}_n$, the representation  is not irreducible anymore. In order to
obtain
irreducible representations of $\hat{sl}_n$, one has to `remove' all elements
$$\sum_{i=1}^s \alpha_{kn_i}^{(i)},\quad k\in {\Bbb Z}.
\tag{4.1}$$
Hence, a KP $\tau$--function is an $\hat{sl}_n$ $\tau$--function if
$$\sum^{s}_{j=1} \frac{\partial \tau}{\partial x^{(j)}_{kn_j}}
= 0,\quad\text{for all $k\in {\Bbb N}$}.\tag{4.2}$$
We will call this
the {\it $[n_1,n_2,\ldots ,n_s]$-th reduced} KP {\it
hierarchy}.
Using the Sato equation (3.13), this
implies the
following  two equivalent conditions:
$$\sum^{s}_{j=1} \frac{\partial V^+(\alpha ,x,z)}{\partial x^{(j)}_{kn_j}} =
V^+(\alpha,x,z)\sum_{j=1}^sz^{kn_j}E_{jj}, $$
$$Q(\alpha)_-=0,\tag{4.3}$$
where $$Q(\alpha)=\sum_{j=1}^sL(\alpha)^{kn_j}C^{(j)}.\tag{4.4}$$

\vskip 10pt
\subheading{\S 5. The string equation and $W_{1+\infty}$ constraints}

\vskip 10pt
\noindent {\it From now on we assume that $\tau$ is any solution of the KP
hierarchy.} In particular, we no longer assume that $\tau_{\alpha}$ is a
polynomial. Of course this means that the corresponding wave functions
$V^{\pm}(\alpha ,x,z)$ will be of a more general nature than before.

Let $L_{-1}$ be given by (2.7a), the string equation is the following
constraint
on $\tau\in F^{(0)}$:
$$L_{-1}\tau=0. \tag{5.1}$$
Using (2.15) we rewrite $L_{-1}$ in terms of operators on $B^{(0)}$:
$$L_{-1}=\sum_{a=1}^s \{
\delta_a x_{n_a}^{(a)}+
{1\over 2n_a}\sum_{p=1}^{n_a-1} p(n_a-p)x_p^{(a)}x_{n_a-p}^{(a)}
+{1\over n_a}\sum_{k=1}^{\infty}(k+n_a)x_{k+n_a}^{(a)}{\partial\over\partial
x_k^{(a)}}\}.
$$
Since $\tau=\sum_{\alpha\in Q}\tau_{\alpha}e^{\alpha}$,
we find that $L_{-1}\tau_{\alpha}=0$ for all $\alpha\in Q$.
Using a calculation similar to the one in [D], one deduces from (5.1)
that
$$N(\alpha)_-=0,\tag{5.2}$$
where

$$N(\alpha):=
\sum_{a=1}^s \{ {1\over n_a}M(\alpha)L(\alpha)^{1-n_a}C^{(a)}(\alpha)
-{{n_a-1}\over 2n_a}L(\alpha)^{-n_a}C^{(a)}(\alpha)\}.\tag{5.3}
$$
Hence $N(\alpha)$ is a differential operator that satisfies
$$[Q(\alpha),N(\alpha)]=1.$$

{\it From now on we assume that $\tau$ is a $\tau$--function of the
$[n_1,n_2,\ldots,n_s]$-th
reduced KP hierarchy, which satisfies the string equation.}
So, we assume that (4.2) and (5.1) holds. Hence,
for all  $\alpha\in \text{supp}\ \tau$ both
$Q(\alpha)$ and
 $N(\alpha)$ are differential operators. Thus, also
$N(\alpha)^pQ(\alpha)^q$ is a differential operator, i.e.,
$$((\sum_{a=1}^s \{ {1\over n_a}M(\alpha)L(\alpha)^{1-n_a}
-{{n_a-1}\over 2n_a}L(\alpha)^{-n_a})^pL(\alpha)^{qn_a}C^{(a)}(\alpha))_-=0
\quad\text{for } p,q\in {\Bbb Z}_+.\tag{5.4}$$
This leads to
\proclaim{Lemma 5.1}
For all $p,q\in {\Bbb Z}_+$ one has
$$\text{Res}_{z=0}dz\sum_{a=1}^s z^{qn_a}({1\over n_a}
z^{{1-n_a}\over 2}{\partial\over \partial z}z^{{1-n_a}\over 2})^p
V^+(\alpha,x,z)E_{aa}^{\phantom{aa}t} V^-(\alpha,x',z) =0.
\tag{5.5}$$
\endproclaim
Taking the $(i,j)$--th coefficient of (5.5) one obtains
\proclaim{Corollary 5.2}
For all $1\le i,j\le s$  and $p,q\in{\Bbb Z}_+$ one has
$$\text{Res}_{z=0}dz\sum_{a=1}^s z^{qn_a}({1\over n_a}
z^{{1-n_a}\over 2}{\partial\over \partial z}z^{{1-n_a}\over 2})^p
\psi^{+(a)}(z)\tau_{\alpha+\delta_i-\delta_a}\otimes
\psi^{-(a)}(z)\tau_{\alpha+\delta_a-\delta_j}=0.
\tag{5.6}
$$
\endproclaim

Let
$$\align
W^{(p+1)}_{q-p}&=\text{Res}_{z=0}dz\sum_{a=1}^s z^{qn_a}({1\over n_a}
z^{{1-n_a}\over 2}{\partial\over \partial z}z^{{1-n_a}\over 2})^p
:\psi^{+(a)}(z)\psi^{-(a)}(z):\\
&=-\sum_{a=1}^s
\hat r (\frac{1}{n_a}t^{n_aq}(t^{\frac{1-n_a}{2}}\frac{\partial}{\partial t}
t^{\frac{1-n_a}{2}})^pe_{aa})
\\
&=\sum_{a=1}^s
\hat r (t^{\frac{n_a-1}{2}}(-\lambda_a^{q}(\frac{\partial}{\partial
\lambda_a})^p)
t^{\frac{1-n_a}{2}})e_{aa})\qquad\text{where}\ \ \ \ \lambda_a=t^{n_a},
\tag{5.7}
\endalign
$$
and
$$-t^{k+\ell}({\partial\over\partial t})^{\ell}e_{ii}
= \sum_{m\in{\Bbb
Z}}-m(m-1)\cdots(m-\ell+1)E^{(ii)}_{-m-k-\frac{1}{2},-m-\frac{1}{2}}.
\tag{5.8}
$$
Using a generalization of an
identity of Date, Jimbo, Kashiwara and Miwa [DJKM3] (see also
 [G], [V]), (5.6) is equivalent to
$$\text{Res}_{y=0}dy
\sum_{a=1}^s \psi^{+(a)}(y)
W^{(p+1)}_{q-p}
\tau_{\alpha+\delta_i-\delta_a}(x)e^{\alpha+\delta_i-\delta_a}
\psi^{-(a)}(y)'\tau_{\alpha+\delta_a-\delta_j}(x')(e^{\alpha+\delta_a-\delta_j})'
=0.
\tag{5.9}
$$

If we ignore the cocycle term for a moment, then it is obvious from (5.7), that
the elements $W^{(p+1)}_k$ are the generators
of the W--algebra $W_{1+\infty}$ [Ra], [KRa] (the cocycle term, however, will
be slightly
different). Upto some  modification of the elements $W^{(p+1)}_0$, one gets
the standard commutation relations of $W_{1+\infty}$, where $c=nI$.

As the next step, we take in (5.9) $x_k^{(i)}=x_k^{(i)\prime}$, for all
$k\in{\Bbb N},\ 1\le i\le s$, we then obtain
$$\cases
\frac{\partial}{\partial
x_1^{(i)}}(\frac{W^{(p+1)}_{q-p}\tau_{\alpha}}{\tau_{\alpha}})=0&
\text{if } i=j,\\
\tau_{\alpha+\delta_i-\delta_j}W^{(p+1)}_{q-p}\tau_{\alpha}=
\tau_{\alpha}W^{(p+1)}_{q-p}\tau_{\alpha+\delta_i-\delta_j}&
\text{if }i\ne j.
\endcases
\tag{5.10}
$$
The last equation means that for all $\alpha,\beta\in \text{supp}\ \tau$ one
has
$$\frac{W^{(p+1)}_{q-p}\tau_{\alpha}}{\tau_{\alpha}}=
\frac{W^{(p+1)}_{q-p}\tau_{\beta}}{\tau_{\beta}}.
\tag{5.11}
$$
Next we divide (5.9) by $\tau_{\alpha}(x)\tau_{\alpha}(x')$, of course
only for $\alpha\in \text{supp}\ \tau$, and use (5.11). Then for all
$\alpha, \beta\in\text{supp}\ \tau$ and $p,q\in {\Bbb Z}_+$ one has
$$\align
\text{Res}_{z=0}dz&
\sum_{a=1}^s \exp(-\sum_{k=1}^{\infty}\frac{z^{-k}}{k}
\frac{\partial}{\partial x_k^{(a)}})
(\frac{W^{(p+1)}_{q-p}\tau_{\beta}(x)}{\tau_{\beta}(x)})
\frac{\psi^{+(a)}(z)\tau_{\alpha+\delta_i-\delta_a}(x)}{\tau_{\alpha}(x)}
e^{\alpha+\delta_i-\delta_a}\times
\\
&\frac{\psi^{-(a)}(z)'\tau_{\alpha+\delta_a-\delta_j}(x')}{\tau_{\alpha}(x')}
(e^{\alpha+\delta_a-\delta_j})'
=0.\endalign
$$
Since one also has the bilinear identity (3.3), we can subtract that part and
thus
obtain the following
\proclaim
{Lemma 5.3}
For all
$\alpha, \beta\in\text{supp}\ \tau$ and $p,q\in {\Bbb Z}_+$ one has
$$\align
\text{Res}_{z=0}dz&
\sum_{a=1}^s \{\exp(-\sum_{k=1}^{\infty}\frac{z^{-k}}{k}
\frac{\partial}{\partial x_k^{(a)}})-1\}
(\frac{W^{(p+1)}_{q-p}\tau_{\beta}(x)}{\tau_{\beta}(x)})\times
\\
&\frac{\psi^{+(a)}(z)\tau_{\alpha+\delta_i-\delta_a}(x)}{\tau_{\alpha}(x)}
e^{\alpha+\delta_i-\delta_a}
\frac{\psi^{-(a)}(z)'\tau_{\alpha+\delta_a-\delta_j}(x')}{\tau_{\alpha}(x')}
(e^{\alpha+\delta_a-\delta_j})'
=0.
\tag{5.12}\endalign
$$
\endproclaim
Define
$$S(\beta,p,q,x,z):=\sum_{a=1}^s
\{\exp(-\sum_{k=1}^{\infty}\frac{z^{-k}}{k}
\frac{\partial}{\partial x_k^{(a)}})-1\}
(\frac{W^{(p+1)}_{q-p}\tau_{\beta}(x)}{\tau_{\beta}(x)})E_{aa}.
$$
Notice that the first equation of (5.10) implies that
$\partial\circ S(\beta,p,q,x,\partial)= S(\beta,p,q,x,\partial)\circ\partial$.
Then viewing (5.12) as the $(i,j)$--th entry of a matrix, (5.12) is
equivalent to
$$\text{Res}_{z=0}dz P^+(\alpha)R^+(\alpha)S(\beta,p,q,x,\partial)e^{x\cdot z}
\ ^t(P^-(\alpha)'R^-(\alpha)'e^{-x'\cdot z})=0.
\tag{5.13}
$$
Now using Lemma 3.1, one deduces
$$(P^+(\alpha)R^+(\alpha)S(\beta,p,q,x,\partial)
R^+(\alpha)^{-1}P^+(\alpha)^{-1})_-=0,
\tag{5.14}
$$
hence
$$P^+(\alpha)S(\beta,p,q,x,\partial)
P^+(\alpha)^{-1}=(P^+(\alpha)S(\beta,p,q,x,\partial)
P^+(\alpha)^{-1})_-=0.
$$
So $S(\beta,p,q,x,\partial)=0$
and therefore
$$\{\exp(-\sum_{k=1}^{\infty}\frac{z^{-k}}{k}
\frac{\partial}{\partial x_k^{(a)}})-1\}
(\frac{W^{(p+1)}_{q-p}\tau_{\beta}(x)}{\tau_{\beta}(x)})=0.
$$
{}From which we conclude that
$$W_{q-p}^{(p+1)}\tau_{\beta}=\text{constant }\tau_{\beta}\quad\text{for all}\
p,q\ge 0.
\tag{5.15}
$$
In order to determine the constants on the right--hand--side of (5.15)
we calculate the Lie brackets
$$[W_{-1}^{(2)},\frac{-1}{q+1}W_{q-p+1}^{(p+1)}]\tau_{\beta}=0.
\tag{5.16}
$$
Notice that the right--hand--side of (5.16) is equal to
$$(W_{p-q}^{(p+1)}+\mu(W_{-1}^{(2)},\frac{-1}{q+1}W_{q-p+1}^{(p+1)}))\tau_{\beta}.$$
Now using (1.2-3), (2.2)  and (5.8) we thus obtain
 the main result, generalizing some of the results of [FKN], see also [AV]:
\proclaim
{Theorem 5.4}
The following two conditions for $\tau\in F^{(0)}$ are equivalent:

\noindent (1) $\tau$ is a $\tau$--function of the $[n_1,n_2,\ldots,n_s]$--th
reduced
$s$--component KP hierarchy which satisfies the string equation (5.1).

\noindent (2) For all $ p,q\ge 0$:
$$(W_{q-p}^{(p+1)}+\delta_{pq}c_{p+1})\tau=0,
\tag{5.17}$$
where
$$c_{p+1}={1\over p+1}\sum_{b=1}^s \sum_{j=1}^{n_b }
({n_b-2j+1\over 2n_b})_{p+1},\tag{5.18}$$
and $(k)_{\ell}=k(k-1)(k-2)\cdots (k-\ell+1)$
\endproclaim
%
\noindent {\bf Aknowledgements.}  I would like to thank Frits Beukers and
Victor Kac for valuable discussions, and
the Mathematical Institute of the University of Utrecht
for the computer and e-mail facilities.

\vskip 20pt
\noindent J.W. van de Leur

\noindent Kievitdwarsstraat 22

\noindent  3514VE Utrecht, The Netherlands

\enddocument